\documentclass[cameraready]{Interspeech}

\title{Scalable Keyword Spotting via Modular Network Expansion}

\author[affiliation={1}]{Viktor}{Khaymonenko}
\author[affiliation={2}]{Dzmitry}{Saladukha}
\author[affiliation={1}]{Aliaksei}{Rak}
\author[affiliation={1}]{Alexander}{Rostov}

\address{
    $^1$ Embedded Voice Input Team, Yandex, Russia \\
    $^2$ Embedded Voice Input Team, Yandex, Belarus
}

\email{khaymonenko@yandex-team.ru, dsoloduha@yandex-team.ru, alexrak@yandex-team.ru, a-rostov@yandex-team.ru}


\usepackage{amsmath}
\usepackage{comment}
\usepackage{multirow}

\begin{document}

\maketitle

\begin{abstract}
Keyword spotting (KWS) models on embedded devices often need to add new keywords after deployment, but updates are difficult when original training data are unavailable and regressions on existing triggers are unacceptable.

At a fixed operating point, our method reduces average new-keyword false reject rate (FRR) from \textbf{6.46\%} to \textbf{4.37\%} versus a parameter-matched separate-model baseline and outperforms parameter-efficient tuning baselines (adapters, LoRA), while using fewer multiply-accumulate operations (MACs) under the same added-parameter budget ($\leq$10k): \textbf{16.34M} vs \textbf{18.45M}/\textbf{20.52M}.

We achieve this via parameter-capped modular expansion: the base network, including batch-normalization statistics and the core classifier, is frozen, and only a lightweight expansion branch with a separate new-keyword head is trained, preserving core logits, shipped outputs, and thresholds for existing keywords.
\end{abstract}

\section{Introduction}
\label{introduction}

Keyword spotting enables hands-free interaction on resource-constrained devices, where small-footprint models must remain accurate and reliable across speakers and acoustic conditions \cite{sainath15b_interspeech,kim21_interspeech,saladukha25_interspeech}. In deployed products, keyword inventories can change over time as new features, locales, or user needs emerge. Updating a fixed-vocabulary KWS model in this setting is difficult: the original training data may be unavailable (e.g., due to privacy or storage constraints), full retraining is compute-intensive, and, more critically, it can introduce regressions on keywords that users already rely on. Open-vocabulary and few-shot approaches reduce the need for retraining, but they often lag behind fixed-vocabulary systems in robustness, especially under noise and for phonetically similar commands. In this work, we focus on \emph{small-footprint embedded KWS} and study \emph{class-incremental keyword expansion}: adding new triggers after deployment under strict update safety constraints.

We target an expansion setting with two practical constraints: the original training data are unavailable during the update, and the updated system must not degrade performance on previously supported keywords. To meet these constraints, we freeze the entire pretrained KWS model, including batch-normalization parameters and statistics, and attach a lightweight trainable expansion path with a separate classifier head for the new keywords. The expansion adds at most 10k parameters to a 150k-parameter base model and is trained only on new-keyword data. Because the base path is unchanged, the logits for the original keyword set are identical to those of the pretrained model for any input, providing a strict non-regression guarantee by construction.

Experiments on Google Speech Commands v2 (GSC) \cite{warden2018speech} demonstrate that this safety guarantee does not prevent effective expansion: under a 10k-parameter budget, modular expansion improves new-keyword detection relative to parameter-matched baselines (including a separate-model deployment baseline, continual-learning methods, and parameter-efficient fine-tuning), while preserving the shipped core detector exactly.

\section{Related work}
\label{sec:related_work}

Keyword spotting has been widely studied due to its importance in always-on voice interfaces. Most deployed systems use a fixed vocabulary and train compact neural classifiers, including CNN-based architectures and streaming-friendly designs such as SVDF models \cite{sainath15b_interspeech, arik17_interspeech, zhuang16_interspeech, alvarez19_icassp}. These models achieve strong accuracy under tight latency and memory constraints, but extending the label set typically requires retraining and careful class balancing, and updates may introduce regressions on previously supported keywords.

To improve flexibility, open-vocabulary and query-based KWS methods have been explored \cite{chen15_icassp, jung25b_interspeech}. These approaches enable detecting user-specified phrases, but in embedded settings they may trade off accuracy and robustness relative to fixed-vocabulary systems, particularly under noise or when distinguishing phonetically similar triggers.

Another direction is fast adaptation via transfer learning and few-shot learning. Metric-based approaches, including prototypical networks and related embedding methods \cite{parnami_2022, huh21_slt}, can add new keywords with limited labeled audio. However, maintaining stable performance on the original keyword set remains challenging when the model is updated, especially when the original training data are unavailable.

Continual learning methods aim to reduce catastrophic forgetting during incremental updates. Regularization-based approaches such as EWC and LwF \cite{kirkpatrick17_nas, li2018} constrain parameter drift, while replay-based methods rehearse past data or generated samples \cite{shin17_nips, rolnick19_nips}. In practice, these methods typically provide soft protection rather than a strict non-regression guarantee, and replay is incompatible with settings where the original training data cannot be stored or reused. Parameter-efficient fine-tuning methods (e.g., adapters or LoRA \cite{houlsby2019_corr, hu2021_corr}) reduce the number of trainable parameters, but they still alter the shared computation path and therefore cannot guarantee unchanged behavior on existing keywords; retaining the original behavior may also require extra inference compute.

Progressive Neural Networks (PNNs) \cite{rusu16_arxiv} address task-incremental learning by freezing previous columns and adding a new column with lateral connections, typically assuming that the task identity (and thus which column to execute) is known at inference time. In contrast, we study \emph{class-incremental expansion} of a \emph{single always-on, fixed-latency} embedded KWS detector, where (i) the original training data are unavailable, (ii) model growth must be strictly bounded, and (iii) the shipped detector’s outputs and threshold-based decision rule for existing keywords must remain \emph{exactly unchanged} for all inputs. Concretely, we freeze the entire deployed network (including batch-normalization statistics) and attach a single parameter-capped ($\leq$10k) expansion branch and a separate head for the new keywords, enabling non-regression by construction.

\section{Methodology}
\label{sec:methodology}

\subsection{Problem setup}
\label{ssec:problem}

Let \(x\) denote an input audio utterance. Let $\mathcal{Y}_1$ denote the set of deployed (core) keywords and $\mathcal{Y}_2$ a disjoint set of new keywords, with $\mathcal{Y}_1 \cap \mathcal{Y}_2 = \varnothing$. We use a single background label $\emptyset$ to denote ``none of the keywords of interest.''

A base model $f_{\mathrm{base}}(x;\theta_{\mathrm{base}})$ is trained on a dataset $D_1$ with labels in $\mathcal{Y}_1 \cup \{\emptyset\}$. We are then given a second dataset $D_2$ with labels in $\mathcal{Y}_2 \cup \{\emptyset\}$ and \emph{no access to} $D_1$ during expansion.

We seek an expanded model $f_{\mathrm{exp}}(x;\theta_{\mathrm{base}},\theta_{\mathrm{exp}})$ that adds a small number of parameters $\theta_{\mathrm{exp}}$, is trained only on $D_2$, and satisfies a strict non-regression property on $\mathcal{Y}_1$ while maximizing performance on $\mathcal{Y}_2$.

\begin{figure}[t]

\begin{minipage}[b]{0.48\linewidth}
  \centering
  \centerline{\includegraphics[width=3.8cm]{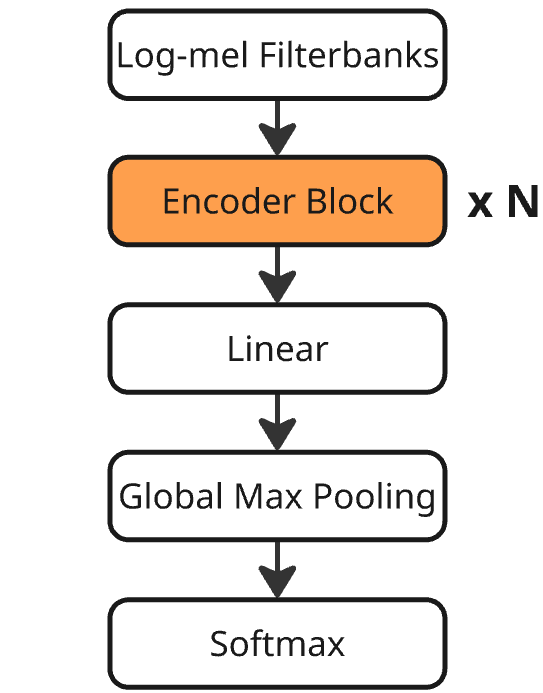}}
  \centerline{(a) Base model}\medskip
\end{minipage}
\hfill
\begin{minipage}[b]{0.48\linewidth}
  \centering
  \centerline{\includegraphics[width=2.3cm]{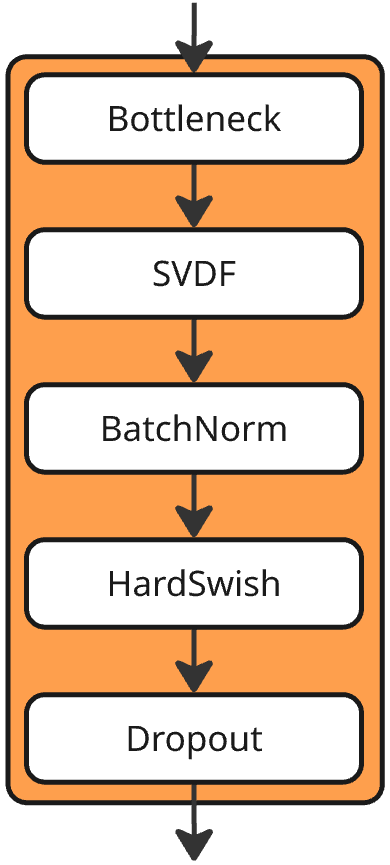}}
  \centerline{(b) Encoder Block}\medskip
\end{minipage}

\caption{Base model and Encoder Block for core-keyword classification.}
\label{fig:architecture}
\end{figure}

\begin{figure}[t]

\begin{minipage}[b]{.48\linewidth}
  \centering
  \centerline{\includegraphics[width=4.0cm]{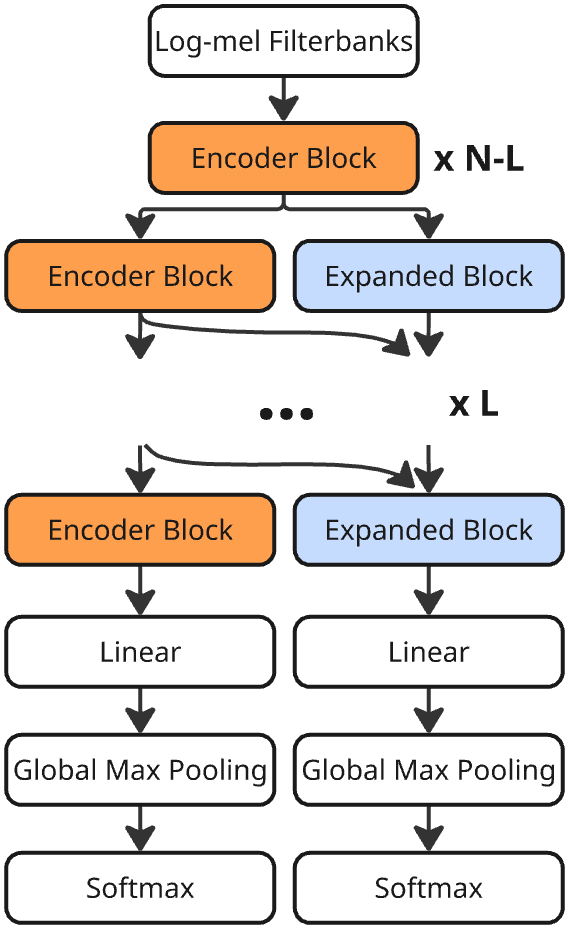}}
  \centerline{(a) Expanded model}\medskip
\end{minipage}
\hfill
\begin{minipage}[b]{0.48\linewidth}
  \centering
  \centerline{\includegraphics[width=3.0cm]{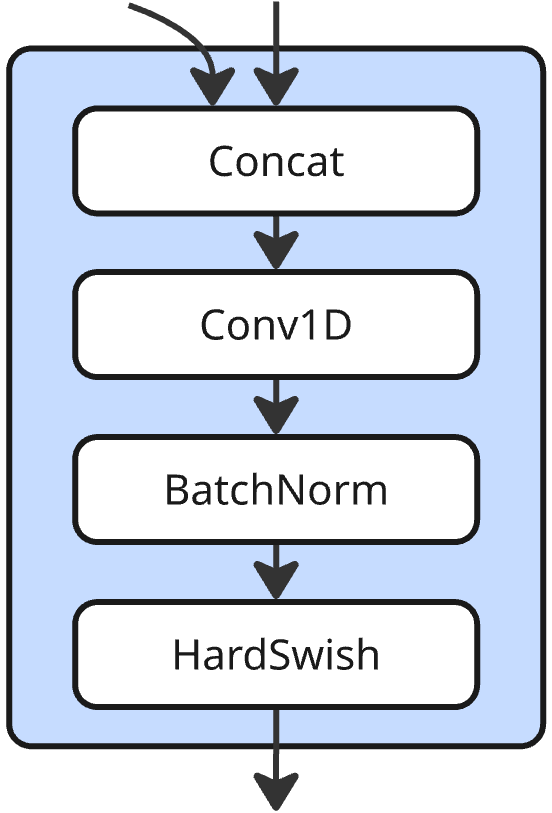}}
 \vspace{1.0cm}
  \centerline{(b) Expanded Block}\medskip
\end{minipage}

\caption{Expanded model. Each Expanded Block taps a frozen Base Block activation, merges it with the previous expanded state, and applies a lightweight residual transform.}

\label{fig:expanded_architecture}
\end{figure}

\subsection{Base architecture}
\label{ssec:base_architecture}

Base network follows a small-footprint SVDF-style architecture~\cite{alvarez19_icassp, nakkiran15_interspeech}. Each Base (Encoder) Block applies SVDF, followed by batch-normalization (BN) \cite{ioffe15_pmlr}, hard-swish~\cite{howard19_iccv}, and dropout~\cite{hinton12_corr}. The encoder output is temporally pooled and passed to an \emph{Core Head} classifier producing logits
$\ell_{\mathrm{core}}^{\mathrm{base}}(x) \in \mathbb{R}^{|\mathcal{Y}_1|+1}$
over $\mathcal{Y}_1 \cup \{\emptyset\}$, followed by softmax. The base model has approximately 150k parameters (Fig.~\ref{fig:architecture}).

\subsection{Modular expansion}
\label{ssec:expansion}

We freeze the entire base path (including BN affine parameters and running statistics) and attach a lightweight, trainable expansion path composed of $L$ Expanded Blocks plus a \emph{New Head} that predicts logits $\ell_{\mathrm{new}}(x)$ over $\mathcal{Y}_2\cup\{\emptyset\}$ (Fig.~\ref{fig:expanded_architecture}). Expanded Blocks tap selected frozen Base Block activations (and optionally the previous Expanded Block output) and apply a small residual transform; the total added parameters (Expanded Blocks + New Head) are capped at 10k (final model $\sim$160k parameters).

Each Expanded Block takes the activation from a frozen Base Block and the previous Expanded Block output, concatenates them along the channel dimension, and applies a lightweight transform consisting of a $1$D convolution (Conv1D over time, implemented as a small-kernel temporal conv), followed by batch-normalization and hard-swish.
The first Expanded Block is a special case: it takes only the frozen encoder activation (no previous expanded output), so the concatenation reduces to an identity.
The block output is then used as input to the next Expanded Block, and the final Expanded Block output is temporally pooled and fed to the New Head.

The expanded model outputs
\begin{align}
p_{\mathrm{core}}(y\mid x) &= \mathrm{softmax}(\ell_{\mathrm{core}}(x)), \quad y\in\mathcal{Y}_1\cup\{\emptyset\},\\
p_{\mathrm{new}}(y\mid x) &= \mathrm{softmax}(\ell_{\mathrm{new}}(x)), \quad y\in\mathcal{Y}_2\cup\{\emptyset\}.
\end{align}
Since the Core Head computation is frozen and does not depend on the expansion path, for all $x$:
\begin{equation}
\begin{aligned}
\ell^{\mathrm{exp}}_{\mathrm{core}}(x) &\equiv \ell^{\mathrm{base}}_{\mathrm{core}}(x), \\
\Rightarrow\quad
p^{\mathrm{exp}}_{\mathrm{core}}(y\mid x) &\equiv p^{\mathrm{base}}_{\mathrm{core}}(y\mid x),
\quad \forall y\in \mathcal{Y}_1\cup\{\emptyset\}.
\end{aligned}
\end{equation}

Training uses only $D_2$ and updates only $\theta_{\mathrm{exp}}$ by minimizing cross-entropy on the New Head:
\begin{equation}
\min_{\theta_{\mathrm{exp}}}\ \ \mathbb{E}_{(x,y)\sim D_2}\left[-\log p_{\mathrm{new}}(y\mid x)\right].
\end{equation}

\begin{table}[t]
	\centering
    \caption{Model size and the worst-case compute cost under the deployed inference pipeline (core-first decision rule) for 1-second utterance.}
	\begin{tabular}{r|c|c||c}
		\multicolumn{1}{c|}{\textbf{Method}} & \textbf{Params (K)} & \textbf{MACs (M)} & \textbf{Rel. MACs} \\
		\toprule\toprule
		Base Model                           & 150                 & 15.08             & $1.00\times$       \\
		Full Finetune                        & 150                 & 15.08             & $1.00\times$       \\
		EWC                                  & 150                 & 15.08             & $1.00\times$       \\
		Head-Only                            & 151                 & 15.16             & $1.01\times$       \\
		Ensemble                             & 160                 & 16.14             & $1.07\times$       \\
		Adapters                             & 160                 & 18.45             & $1.22\times$       \\
		LoRA                                 & 160                 & 20.52             & $1.36\times$       \\
		\hline\hline
		Proposed                             & 160                 & 16.34             & $1.08\times$       \\
	\end{tabular}
			
	\vspace{1mm}
	\footnotesize
    \label{tab:macs_onecol}
\end{table}

\subsection{Inference: core-first decision rule}
\label{ssec:inference_rule}

In a deployed KWS system, preserving the shipped behavior means preserving the
\emph{decision rule} of the original detector for the core label set
$\mathcal{Y}_1\cup\{\emptyset\}$, not only its logits.
We therefore use a core-first decision rule.

Let $s^{\mathrm{core}}_y(x)=p_{\mathrm{core}}(y\mid x)$ for $y\in\mathcal{Y}_1$
and let $\tau^{\mathrm{core}}_y$ denote the \emph{same} per-keyword thresholds
used by the shipped model.
The core detector outputs
\begin{equation}
\begin{aligned}
y_{core}^{*}(x) &= \arg\max_{y\in\mathcal{Y}_1} s^{\mathrm{core}}_{y}(x),\\
\hat{y}_{\mathrm{core}}(x) &=
\begin{cases}
y_{core}^{*}(x), & s^{\mathrm{core}}_{y_{core}^{*}(x)}(x)>\tau^{\mathrm{core}}_{y_{core}^{*}(x)},\\
\emptyset, & \text{otherwise}.
\end{cases}
\end{aligned}
\end{equation}

Only if the core detector rejects (outputs $\emptyset$) do we evaluate the new head.
Let $s^{\mathrm{new}}_y(x)=p_{\mathrm{new}}(y\mid x)$ for $y\in\mathcal{Y}_2$
with thresholds $\tau^{\mathrm{new}}_y$. The expanded system outputs
\begin{equation}
\begin{aligned}
y_{\mathrm{new}}^{*}(x) &= \arg\max_{y\in\mathcal{Y}_2} s^{\mathrm{new}}_{y}(x),\\
\hat{y}(x)&=
\begin{cases}
\hat{y}_{\mathrm{core}}(x), & \hat{y}_{\mathrm{core}}(x)\neq\emptyset,\\
y_{\mathrm{new}}^{*}(x), &
s^{\mathrm{new}}_{y_{\mathrm{new}}^{*}(x)}(x) > \tau^{\mathrm{new}}_{y_{\mathrm{new}}^{*}(x)}, \\
\emptyset, & \text{otherwise}.
\end{cases}
\end{aligned}
\end{equation}

\subsection{Baselines}
\label{ssec:baselines}

We compare against four baseline families that differ in whether they modify the
shipped core detector and whether they can provide a strict non-regression
guarantee. Unless stated otherwise, adaptation uses only $D_2$ and all methods
that add capacity are constrained to $\leq$10k new parameters. For completeness,
Table~\ref{tab:macs_onecol} reports the parameter count and worst-case compute
cost in multiply-accumulate operations (MACs) for each method under the deployed core-first inference pipeline; we
use these values to budget-match baselines and to quantify inference overhead.

\textit{Unconstrained model updates.}
\textbf{Full Finetune} continues training the base model on $D_2$ by updating
all parameters with cross-entropy over $\mathcal{Y}_2\cup\{\emptyset\}$.
This serves as a reference for fitting the new keywords, but it generally
introduces catastrophic forgetting on $\mathcal{Y}_1$.

\textit{Core-invariant expansions (safe by construction).}
\textbf{Head-Only} trains only a new classifier head on top of the frozen base
representation (our method with $L{=}0$). \textbf{Ensemble} trains an independent
$\sim$10k-parameter model on $D_2$ and combines it with the shipped detector using
the same core-first decision rule at inference. Both approaches preserve the core
detector exactly, but Head-only often lacks capacity, while Ensemble does not
reuse the base model’s intermediate features.

\textit{Parameter-efficient tuning with a forked branch.}
\textbf{Adapters} and \textbf{LoRA} introduce trainable modules in a parallel
top-$L$ branch that feeds the New Head, while the original Core Head continues to
use the frozen base path. We choose $L$ (and LoRA ranks) to match the additional
inference compute of our expansion under the same 10k-parameter budget.

\textit{Relaxed-constraint continual learning.}
\textbf{EWC} is included for context: it regularizes fine-tuning using Fisher
information estimated from $D_1$ and therefore does not strictly satisfy the
no-access-to-$D_1$ setting; we use $\lambda=5\times 10^{5}$ in all experiments.

\label{experiments}

\begin{table*}[t]
\centering
\caption{Macro FRR (\%) on GSC test at thresholds calibrated to 1\% FAR on CV,
evaluated with the core-first decision rule.}
\begin{tabular}{r|ccccc||c}
           & \{left, right\} & \{on, off\} & \{stop, go\} & \{up, down\} & \{yes, no\} & \textbf{Avg.}\\
\toprule\toprule
\multicolumn{7}{c}{\textbf{Core Phrases}}\\
\bottomrule
Base Model & \textbf{2.37}\textit{(±0.24)}& \textbf{2.42}\textit{(±0.31)}& 3.29\textit{(±0.55)}& 2.63\textit{(±0.28)}& \textbf{2.83}\textit{(±0.12)}& \textbf{2.71}\textit{(±0.12)}\\
Full Finetune & 82.94\textit{(±2.64)}& 79.68\textit{(±7.79)}& 67.94\textit{(±6.30)}& 54.33\textit{(±13.55)}& 60.52\textit{(±5.81)}& 69.08\textit{(±1.98)}\\
EWC        & 7.07\textit{(±0.62)}& 5.02\textit{(±0.32)}& 3.06\textit{(±0.53)}& 2.63\textit{(±0.24)}& 4.56\textit{(±0.16)}& 4.47\textit{(±0.18)}\\
\toprule\toprule
\multicolumn{7}{c}{\textbf{New Phrases}}\\
\bottomrule
Full Finetune & 2.86\textit{(±0.48)}& 2.14\textit{(±0.46)}& 1.60\textit{(±0.50)}& 3.33\textit{(±0.70)}& 1.82\textit{(±0.47)}& 2.35\textit{(±0.27)}\\
Head-Only& 42.55\textit{(±16.44)}& 28.11\textit{(±11.68)}& 18.61\textit{(±10.88)}& 12.13\textit{(±2.43)}& 10.11\textit{(±2.20)}&22.30\textit{(±4.27)}\\
Ensemble   & 8.77\textit{(±2.00)}& 3.81\textit{(±0.99)}& 6.68\textit{(±1.91)}& 5.82\textit{(±0.73)}& 7.21\textit{(±1.17)}& 6.46\textit{(±0.42)}\\
EWC        & \textbf{5.44}\textit{(±0.40)}& 6.97\textit{(±1.08)}& 5.96\textit{(±0.57)}& 12.28\textit{(±1.89)}& 9.09\textit{(±1.18)}& 7.95\textit{(±0.64)}\\
Adapters       & 13.38\textit{(±1.68)}& 8.95\textit{(±1.61)}& 6.68\textit{(±2.01)}& 5.74\textit{(±0.65)}& 5.52\textit{(±0.80)}& 8.05\textit{(±0.35)}\\
LoRA           & 9.88\textit{(±1.36)}& 5.84\textit{(±1.28)}& 6.61\textit{(±2.11)}& 5.12\textit{(±1.78)}& 4.61\textit{(±0.86)}& 6.41\textit{(±0.66)}\\
\hline\hline
Proposed   & 6.69\textit{(±1.04)}& \textbf{2.68}\textit{(±0.39)}& \textbf{4.56}\textit{(±0.86)}& \textbf{4.57}\textit{(±0.73)}& \textbf{3.36}\textit{(±0.38)}& \textbf{4.37}\textit{(±0.33)}\\
\end{tabular}
\label{tab:results_combined}
\end{table*}

\begin{figure}[t]

\begin{minipage}[t]{1.0\linewidth}
  \centering
  \centerline{\includegraphics[width=8.5cm]{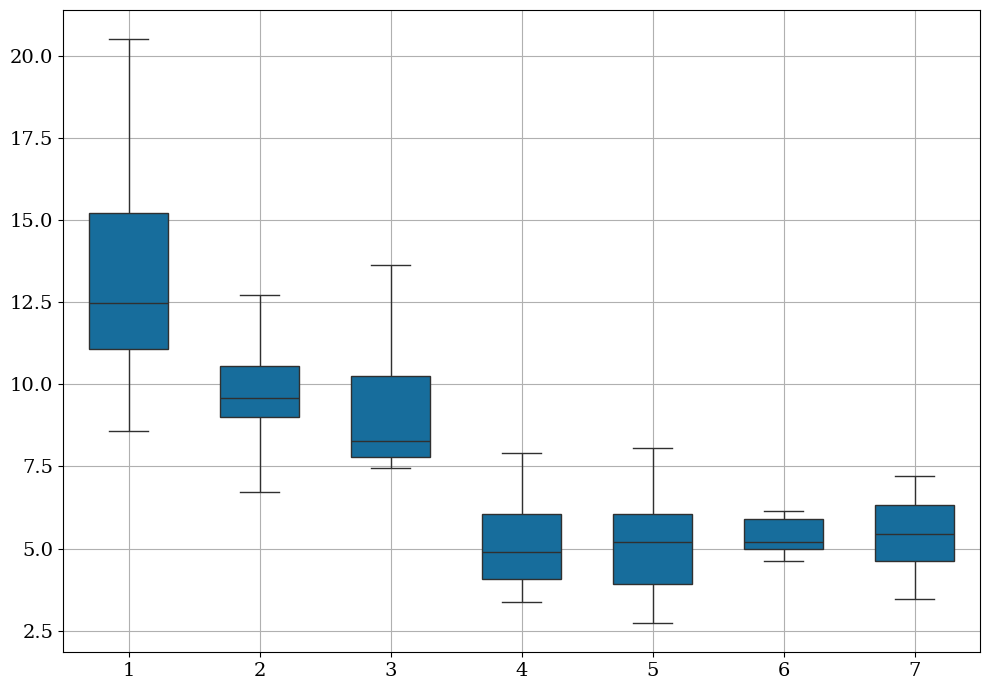}}
\end{minipage}
\caption{New-keyword FRR (\%) vs. Expanded Blocks $L$ for $\mathcal{Y}_2=\{\textit{left,right}\}$.}

\label{fig:shared_blocks_comparison}
\end{figure}

\section{Experiments}
\label{sec:experiments}

\subsection{Datasets and expansion tasks}
Experiments are conducted on Google Speech Commands v2 (GSC) using the official train/validation/test splits. The target vocabulary is the standard set of 10 commands
\{\textit{left, right, on, off, stop, go, up, down, yes, no}\}.
All other words in GSC, together with silence segments, are mapped to the background label $\emptyset$.

Keyword expansion is evaluated through five held-out pairs to reduce sensitivity to any single keyword choice and to cover a range of confusability patterns. For each task, one command pair is designated as the new set $\mathcal{Y}_2$ and the remaining eight commands form the deployed core set $\mathcal{Y}_1$. The held-out pairs are
\{left, right\}, \{on, off\}, \{stop, go\}, \{up, down\}, and \{yes, no\}.
A base model is trained on $D_1$ (GSC train restricted to $\mathcal{Y}_1 \cup \{\emptyset\}$). The expansion stage then uses only $D_2$ (GSC train restricted to $\mathcal{Y}_2 \cup \{\emptyset\}$), with no access to the original $D_1$ audio.

To obtain a less biased negative distribution than GSC itself, Mozilla Common Voice v17 (CV) \cite{ardila-etal-2020-common} is used as a negative-only corpus. In all reported results, positives come from the GSC \emph{test} split, while negatives for FAR calibration come from the CV \emph{test} split.

\subsection{Features and training setup}
Audio is 16\,kHz mono. Input features are 40-bin log-mel filterbanks computed with a 25\,ms window and 10\,ms hop. To improve robustness to channel and speaking-style variability, training applies SpecAugment \cite{park19e_interspeech} directly on the log-mel features (time masking and frequency masking).

Optimization uses Adam \cite{kingma14_iclr} with a cosine learning-rate schedule \cite{loshchilov17_iclr}. Base Models are trained for 40 epochs on $D_1$. Expansion training runs for 10 epochs on $D_2$. For modular methods (including the proposed one), only the newly added parameters are updated during expansion; the base path is kept frozen, including batch-normalization affine parameters and running statistics. Methods that introduce new capacity are constrained to an added-parameter budget of $\leq$10k.

\subsection{Metric, thresholds, and confidence intervals}
Performance is reported as macro false reject rate (FRR) at a fixed false accept rate (FAR) operating point. For each keyword $y$, a threshold $\tau_y$ is chosen to match 1\% FAR on the CV test partition, using the corresponding keyword score (Core Head for $y\in\mathcal{Y}_1$, New Head for $y\in\mathcal{Y}_2$).

Given calibrated thresholds, FRR is computed on the GSC test split. Reported values are macro-averaged across the keywords in the evaluated set (core or new).

Each (method, keyword-pair) configuration is repeated with 8 independent runs (different random seeds). Table~\ref{tab:results_combined} reports the sample mean and a 95\% confidence interval estimated via a $t$-statistic:
$\bar{x}\ \pm\ t_{0.975,7}\, s/\sqrt{8}$,
where $s$ is the sample standard deviation across runs.

\subsection{Comparative analysis}
Table~\ref{tab:results_combined} reports macro FRR (\%) on GSC positives with
thresholds calibrated to 1\% FAR on CV negatives (95\% confidence intervals in
parentheses).

The Base Model achieves low FRR on the core vocabulary (from 2.37\% to 3.29\%,
2.71\% on average). The \emph{Core Phrases} section includes only methods that
can change the shipped detector (Full Fine-tuning and EWC). Adapters, LoRA,
Ensemble, and the proposed method freeze the core path, so their core results
match the Base Model exactly and are omitted to avoid repeating identical
numbers.

Full Fine-tuning fits the new phrases well (2.35\% average FRR) but catastrophically
regresses the core vocabulary, increasing average core FRR from \textbf{2.71\%}
to \textbf{69.08\%}. EWC mitigates forgetting but still increases core FRR to
\textbf{4.47\%}; on new phrases it reaches \textbf{7.95\%} on average (best case
5.44\% on \{left, right\}). We include EWC for context, although it relies on
information derived from $D_1$ (or precomputed Fisher statistics) and thus
violates the strict no-$D_1$ setting.

Among methods that satisfy the deployment constraints, Head-Only performs poorly
on new keywords (\textbf{22.30\%} average FRR). Parameter-efficient baselines
improve substantially, with Adapters at \textbf{8.05\%} and LoRA at
\textbf{6.41\%}. Ensemble is a strong deployment baseline that adds an
independent $\sim$10k-parameter model for $\mathcal{Y}_2$, achieving
\textbf{6.46\%}. Modular expansion further improves new-keyword detection under
the same parameter budget, reducing average FRR to \textbf{4.37\%} (an absolute
gain of \textbf{2.09} points).

\subsection{Ablation Study: Expansion Depth}

We analyze the impact of feature reuse depth in Figure \ref{fig:shared_blocks_comparison}, which plots new-keyword FRR against the number of tapped blocks in the expansion path. Tapping only shallow layers (1–3 blocks) yields high error rates, suggesting insufficient semantic abstraction. Conversely, extending the expansion to four blocks minimizes the median FRR. Beyond this point, performance plateaus or slightly degrades, indicating that the deepest layers of the base network are overly specialized to the core vocabulary $Y_1$ and provide limited transfer utility. Consequently, we adopt the four-block expansion as the optimal configuration for all reported comparisons.
\section{Conclusion}
\label{sec:conclusion}

On-device KWS systems must evolve as products add features and languages, yet
updates are challenging when the original training audio cannot be reused and
regressions on existing keywords are unacceptable.

We presented a modular expansion mechanism that adds new keywords without
touching the deployed detector. The full base network is kept frozen (including
batch-normalization statistics), and new functionality is introduced through a
small attached branch and a separate new-keyword head (up to 10k parameters),
keeping the core detector’s logits and threshold behavior exactly unchanged.

On Google Speech Commands (thresholds calibrated at 1\% FAR using Common Voice),
the expanded model improves average new-keyword FRR from \textbf{6.46\%} to
\textbf{4.37\%} versus a parameter-matched separate-model baseline and also
outperforms adapter- and LoRA-based tuning. It further uses fewer worst-case
MACs under the deployed core-first pipeline (\textbf{16.34M}) than adapters
(\textbf{18.45M}) and LoRA (\textbf{20.52M}) at the same added-parameter budget. Together, these results
support modular expansion as a practical way to scale embedded KWS vocabularies
while preserving the shipped user experience.

Multi-stage expansion, where new keyword sets are added sequentially, is a
natural direction for future work, especially under tight parameter and latency budgets.

\section{Generative AI Use Disclosure}
We used ChatGPT (OpenAI; GPT-5.2) to assist with English editing and polishing of the manuscript. All technical content, experiments, and conclusions were produced and verified by the authors.

\bibliographystyle{IEEEtran}
\bibliography{mybib}

\end{document}